\documentclass[journal]{IEEEtran}

\usepackage{cite}
\usepackage{amsmath,amssymb,amsfonts,amsthm}
\usepackage{algorithmic}
\usepackage{graphicx}
\usepackage{textcomp}
\usepackage{url}
\usepackage[format=hang]{caption}

\usepackage[utf8]{inputenc}   
\usepackage[T1]{fontenc}  
   
\begin{document}

\title{Multi-User Visible Light Communications: State-of-the-Art and Future Directions}

\author{Saad~Al-Ahmadi, Omar~Maraqa, Murat~Uysal, \IEEEmembership{(Senior~Member,~IEEE)} and Sadiq~M.~Sait, ~\IEEEmembership{(Senior~Member,~IEEE)}
%

\thanks{© 2018 IEEE. Personal use of this material is permitted. Permission from IEEE must be obtained for all other uses, in any current or future media, including reprinting/republishing this material for advertising or promotional purposes, creating new collective works, for resale or redistribution to servers or lists, or reuse of any copyrighted component of this work in other works. Published in IEEE Access with DOI: 10.1109/ACCESS.2018.2879885.}}
\maketitle

\begin{abstract}
Visible light communication (VLC) builds upon the dual use of existing lighting infrastructure for wireless data transmission. VLC has recently gained interest as cost-effective, secure, and energy-efficient wireless access technology particularly for indoor user-dense environments. While initial studies in this area are mainly limited to single-user point-to-point links, more recent efforts have focused on multi-user VLC systems in an effort to transform VLC into a scalable and fully networked wireless technology. In this paper, we provide a comprehensive overview of multi-user VLC systems discussing the recent advances on multi-user precoding, multiple access, resource allocation, and mobility management. We further provide possible directions of future research in this emerging topic.
\end{abstract}


\section{Introduction}
\label{sec:introduction}
In the early 80's and 90's, the second generation (2G) cellular systems such as GSM and IS-95 were introduced and operated at frequency bands near 900 MHz and 1900 MHz with support of only low data-rate services such as voice and short message service (SMS). The choice of the aforementioned bands was motivated by the favorable propagation conditions at these bands that allow reasonable coverage for outdoor to indoor communications. The third generation (3G) and fourth generation (4G) standards operated at similar frequency bands below 6 GHz. In these standards, advanced physical layer solutions such as adaptive modulation, turbo coding, single-user and multi-user multiple-input multiple-output (MIMO) transmission and reception schemes, orthogonal frequency  division multiplexing (OFDM), coordinated multi-point (CoMP) transmission schemes, and carrier aggregation~\cite{4g-book1, 4gj1} were adopted to support high date rates in the order of tens of Megabits per second. 

The recent surge in mobile data use and emerging new applications in fifth generation (5G) networks require peak data rates in the order of Gigabits per second. This motivated the move from the current congested radio bands to millimeter wave (mm-wave) band~\cite{mm-wave1}. For example, the recently licensed 28 GHz and 38 GHz bands offer more than one GHz of bandwidth which allows service providers to expand the transmission bandwidth to  more than the 20 MHz used in current 4G networks~\cite{mm-wave1}. The effect of the higher path loss due to higher frequencies, compared to the lower bands, is accommodated through the deployment of dense small cells. The spectral efficiency in 5G and beyond-5G networks will be further enhanced through the use of large-scale or massive MIMO, adaptive beamforming, and non-orthogonal multiple access (NOMA) schemes~\cite{5gj1,5gj2,5gj3}.               

Further need for higher data rates, improved security, and increased energy efficiency has geared attention towards the optical spectrum which offers abundant unregulated bandwidth, low interference, and  secure and cost-effective communication as compared to radio frequency (RF) bands~\cite{optical-book1}. This has spurred the interest in visible light communications (VLC) as a potential candidate for complementing and off-loading RF communication systems for various indoor user-dense scenarios such as homes, office rooms, conference and exhibition halls, airplanes and train cabins. VLC is based on the principle of modulating light emitting diodes (LEDs) without any adverse effects on the human eye and illumination levels. This gives an opportunity to exploit the existing illumination infrastructure for wireless communication purposes. According to a recent market report~\cite{Li-Fi-Market-size}, VLC market size is anticipated to reach 75 Billion US\$ by 2023. In line with such an economic potential, the international standardization efforts were already initiated both by IEEE and ITU, see IEEE 802.15.13 Task Group~\cite{Multi-Gigabits-OWC}, IEEE 802.11.bb~\cite{IEEE-Project-802.11bb} and ITU-T G.vlc~\cite{ITU-T-G.vlc}. In parallel, the first generation of VLC modems are already available from several vendors such as Philips, PureLiFi, OledComm, Velmenni, and VLNComm.

In line with academic and industrial attention on VLC, there has been a growing literature on VLC, see e.g., relevant surveys and monographs~\cite{optical-book1, vlc-book-2018, ow-surv-1, ow-surv-2,ow-surv-3,vlc-surv-1,vlc-surv-2,vlc-surv-3}. Most research efforts have however mostly focused on point-to-point single-user scenarios so far. In an effort to transform VLC into a multi-user, scalable, and fully networked wireless technology, more recent efforts have addressed multi-user VLC networking. Some initial survey papers on multiple access in VLC systems have already appeared in~\cite{vlc-multiuser-surv-1, vlc-mutiuser-surv-2,vlc-mutiuser-surv-3}. However, a large volume of work, beyond the covered literature in~\cite{vlc-multiuser-surv-1} has been reported and the scopes of~\cite{vlc-mutiuser-surv-2,vlc-mutiuser-surv-3} were mainly limited to optical NOMA and multiple access schemes in VLC systems, respectively. The main objective of our paper is to provide a more comprehensive overview of multi-user VLC systems covering multi-user MIMO (MU-MIMO) and beamforming, recent advances in multiple access schemes particularly on NOMA, and mobility management. 

The structure of this paper is as follows: Section~\ref{section: Background and Terminology} summarizes the related multi-user communication schemes and the associated terminology in RF communications. In Section~\ref{section: Single-User Visible Light Communication Systems}, we provide fundamentals on single-user VLC systems and present a brief overview of latest advances. In Section~\ref{section: Multi-user Visible Light Communication Systems} we turn our attention on multi-user VLC systems which is the main focus of this paper. In Section~\ref{section: Conclusion and Directions for Future Research}, we provide conclusions and list some of the possible directions of future research in multi-user VLC communications.

\section{Background and Terminology} 
\label{section: Background and Terminology}
Most of the existing work on multi-user VLC schemes were adopted from or inspired by the multi-user schemes that were originally introduced in the RF literature. Therefore, this section is intended to summarize the related multi-user communication schemes and the associated terminology in RF communications where we overview the relevant basics such as multi-user MIMO and precoding, CoMP, and orthogonal and non-orthogonal multiple access schemes.  

\subsection{Multi-user MIMO and CoMP Schemes in RF Systems} 

 Multi-user MIMO systems have gained interest over the last decade as one of the potential enablers of high data-rate communications~\cite{mu-mimo-j1} and are already adopted as a downlink transmission scheme in LTE-Advanced standard~\cite{4g-book1}. In general, multi-user MIMO systems refer to the communication among a set of wireless terminals that are equipped with multiple antennas. A common example is the uplink and downlink transmissions in cellular networks where a multi-antenna base station (BS) receives from or transmits to single-antenna or multi-antenna user terminals. From an information theoretic perspective, the MIMO uplink and downlink scenarios, in a single-cell with a central processing unit, can be modeled as a vector multiple-access channel (MAC) and a vector broadcast channel (BC), respectively.
 
 In the following, we present a summary of the main optimal and sub-optimal communication schemes for MAC and BC channels. Further elaboration on MU-MIMO schemes can be found in~\cite{mu-mimo-book1, mu-mimo-book2,mu-mimo-j2, mu-mimo-survey}. It is well known that the sum rate capacity of a scalar Gaussian MAC channel can be achieved by joint maximum likelihood detection or successive interference cancellation (SIC) where the stronger user can decode and subtract the signal of the weaker user(s) and then decode its own signal. In a fading scenario with perfect channel state information at the transmitter (CSIT), the sum rate ergodic capacity can be achieved by allocating all the power to the strongest or best user at each channel state~\cite{mu-mimo-book1}. The sum rate capacity of additive white Gaussian noise (AWGN) vector MAC channels can be also achieved by using SIC detection and iterative water-filling~\cite{mac-j1}. This extends to MIMO MAC fading channels with perfect CSIT with joint space and time iterative water-filling~\cite{mac-j1} and the ergodic capacity can be obtained by averaging over the channel stationary statistics.  

On the other hand, the sum rate capacity of scalar Gaussian BC channels can be achieved by the use of superposition coding (SC) at the transmitter and SIC at the receivers~\cite{mu-mimo-book1}. However, a vector BC channel is no more a simple degraded BC (i.e., the users may not be simply ranked by their channel gains) and the capacity-achieving scheme is dirty paper coding (DPC)~\cite{broadcast-j1}. In a fading scenario with perfect CSIT, the sum rate ergodic capacity for a scalar BC can be achieved by transmitting to the strongest or best user at each time (opportunistic user scheduling or multi-user diversity) while the sum rate capacity for vector BC channels can be achieved also by DPC. The capacity of a vector BC with individual (per-antenna) power constraints and perfect CSIT was addressed by Yu and Lan  using the uplink-downlink duality and efficient numerical optimization techniques to obtain the achievable rate region~\cite{broadcast-per-antenna}. The capacity of vector BC fading channels with partial channel state information (CSI) is still an open problem.  

To reduce the computational complexity of the optimal DPC scheme, the zero-forcing DPC (ZF-DPC) scheme, where a part of the multi-user interference is mitigated through the QR decomposition and the other part through successive dirty-paper encoding, was proposed in \cite{broadcast-j1}. Moreover, linear precoding techniques, such as zero-forcing (ZF) and block-diagonalization (BD) were introduced  by Spencer et al. and Peel et al.~\cite{broadcast-linj1,broadcast-linj2}. In BD, the multi-user interference is eliminated by selecting the beamforming vectors for the desired user that lies in the null space of the matrix containing the channel matrices of the other interfering users.

 In multi-cell scenarios, practical considerations have led to the introduction of multi-cell cooperation schemes, known as network MIMO or CoMP schemes, to eliminate or reduce the inter-cell interference (ICI)~\cite{comp-1, comp-2}. These schemes are based upon various degrees of coordination. In joint transmission, all the BSs or the access points (APs) exchange both the user data and CSI while, in selective joint transmission, only the serving BS or a cluster of BSs transmit to their users. Another alternative is coordinated beamforming where only the CSI is shared among the BSs to allow reducing ICI through appropriate beamforming techniques. The latter two schemes were introduced to reduce the complexity and overhead of the full joint cooperation scheme.

\subsection{Multiple Access in RF Systems} 
\label{subsection: NOMA}

In 2G cellular systems, conventional schemes such as time division multiple access (TDMA) and frequency division multiple access (FDMA) were used to support multiple users. The 3G cellular systems adopted code division multiple access (CDMA) where the set of users employ spreading or signature sequences to allow orthogonal multiple access while sharing the same time and frequency. In CDMA systems, the detection can be either based on single-user detection where the other users interference is treated as additional noise or multi-user detection where all the users data are jointly detected using optimal maximum likelihood (ML) detection or linear ZF or minimum mean square error (MMSE) detection schemes. In 4G networks, orthogonal division multiple access (OFDMA) was used where the sub-carriers are assigned to the users in an interleaved or random manner.

5G networks and beyond need to guarantee services for a large number of high data-rate users that share the same resources. Non-orthogonal multiple access (NOMA) is a multiple access candidate scheme for beyond-5G networks~\cite{ding2015cooperative} which can be applied in both MAC and BC channels. In contrast to orthogonal multiple access schemes such as TDMA, CDMA, FDMA, NOMA allows multiple users to allocate same resources in the same frequency, time, and space dimensions. There are three major versions of NOMA, namely, the power-domain NOMA, the code-domain NOMA, and NOMA multiplexing in multiple domains (i.e., the power domain, the code domain, and the spatial domain)~\cite{choi2017noma}. However, the widely adopted one is the power-domain NOMA where the channel gain difference is exploited at the receiver to separate the multiplexed users signals. 

The advantages of NOMA schemes includes offering higher cell-edge throughput, improving the spectral efficiency, achieving user fairness, and attaining low transmission latency. This is at the cost of:  (i) increasing the receiver implementation complexity, and, (ii) increasing the mutual interference among the users sharing the same resources~\cite{dai2015non,islam2017power}. User-grouping or user-clustering techniques are employed to decrease the receiver implementation complexity (i.e., the first disadvantage of NOMA schemes) and multi-user transmission-reception techniques such as SC at the transmitter and SIC at the receiver are employed to reduce the mutual interference among the users (i.e., the second disadvantage of NOMA schemes). 

\begin{figure*}[!t]
\centering
\includegraphics[width=300pt]{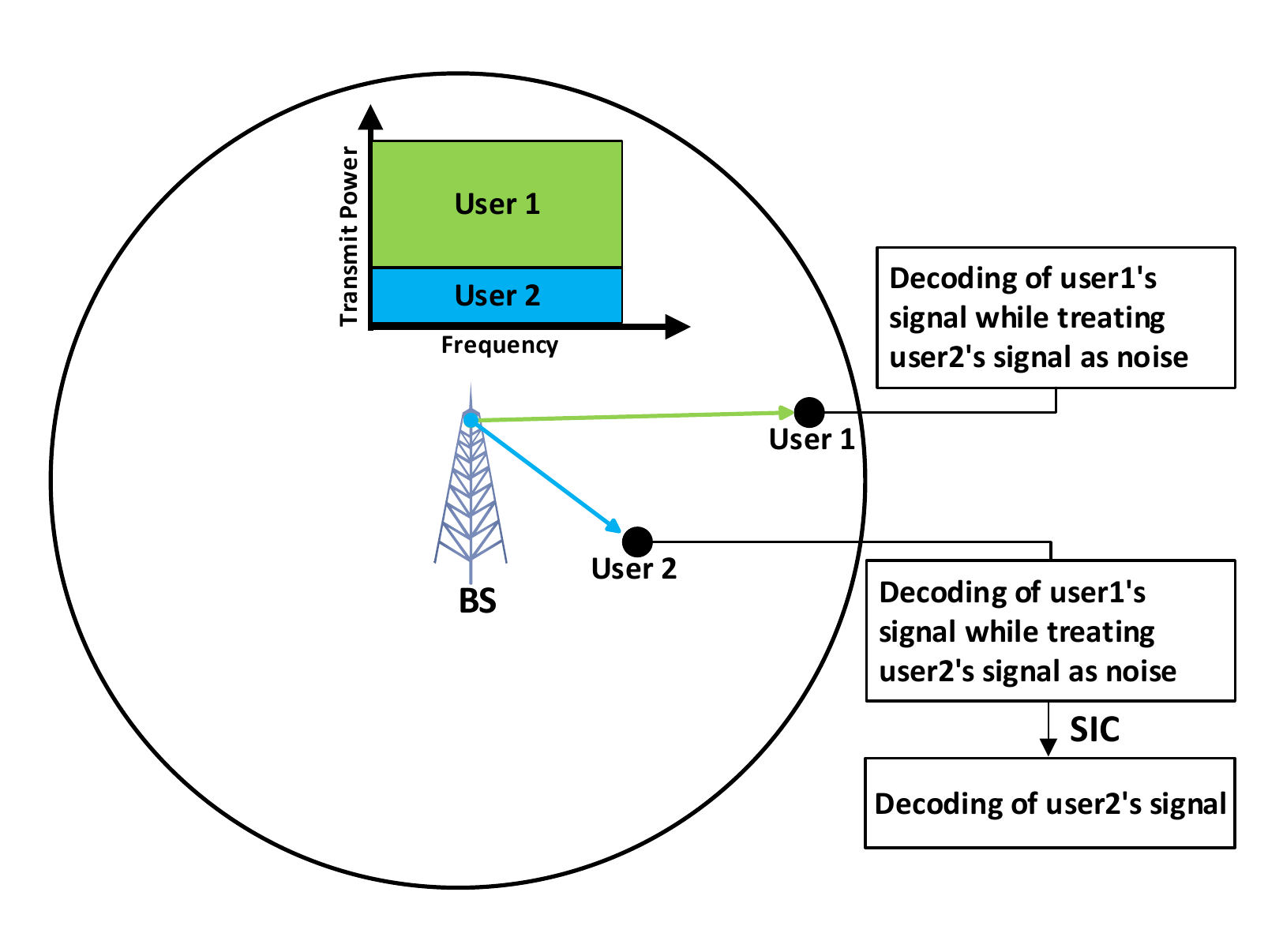}
\caption{An illustration of a two-user downlink power-domain NOMA scheme \cite{5gj3}.}
\label{fig:NOMA_2users}
\end{figure*}

To explain the concept of the power-domain NOMA scheme with an illustration, we consider a two-users downlink NOMA scheme in Fig.~\ref{fig:NOMA_2users}. User 1 (weak user) is farther away from the BS with weak channel gain and User 2 (strong user) is close to the BS with strong channel gain. At the transmitter, both signals for the weak and the strong users are superimposed upon each other with different power allocation for each user. Intuitively, the transmitter allocates more power for the weak user as it has a larger path loss, as compared to the strong user (note that the size of the rectangles for User 1 and User 2 are different indicating the different power allocated for each user). At the strong user receiver, the signal of the weak user has high signal to noise ratio (SNR) which implies that the strong user can successfully decode and subtract the weak user's signal before decoding its own signal (i.e., performing SIC). On the other hand, at the weak user receiver, the strong user signal is considered as noise as its transmission power is lower than the weak user signal. Subsequently, the weak user can decode its signal directly without SIC~\cite{choi2017noma}.

With its promising performance, NOMA was considered as a candidate in various standardization activities. For LTE-A systems (3rd Generation Partnership Project (3GPP) release 13), NOMA has been considered under the name of multi-user superposition transmission (MUST). LTE-A/LTE-A Pro (3GPP releases 13 and 14) incorporates a new category of user terminals that have interference cancellation capability known as network assisted interference cancellation and suppression (NAICS). Furthermore, in the initial studies of 5G new radio (NR), the standardization body recognized that at least uplink NOMA should be considered especially for massive machine type communications. NOMA schemes are also proposed in the context of 3GPP releases 15 and 16~\cite{3Gpp-proposal-release-15,3Gpp-release-16}.

\begin{figure*}[!t]
\centering
\includegraphics[width=450pt]{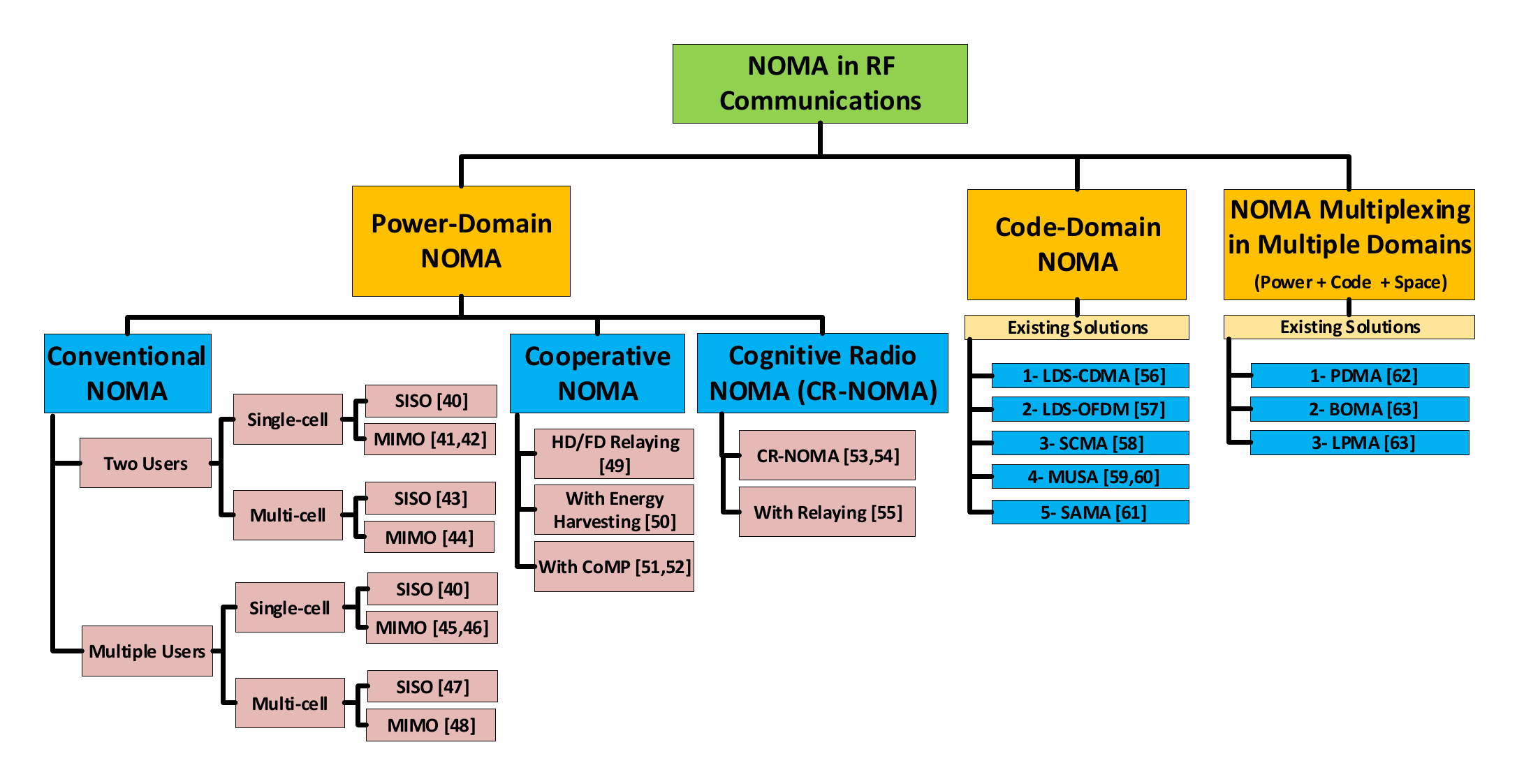}
\caption{A taxonomy of existing NOMA schemes in the RF literature ("SISO": "single input single output", "LDS-CDMA": "low-density spreading CDMA", "LDS-OFDM": "low-density spreading OFDM", "SCMA":"sparse code multiple access", "MUSA": "multi-user sharing access", "SAMA": "successive interference cancellation amenable multiple access", "PDMA": "pattern division multiple access", "BOMA": "building block sparse-constellation based orthogonal multiple access", and "LPMA": "lattice partition multiple access").}
\label{fig:NOMA_in_RF}
\end{figure*}

\nocite{NOMA-Tree-40,NOMA-Tree-41,NOMA-Tree-42,NOMA-Tree-43,NOMA-Tree-44,NOMA-Tree-45,NOMA-Tree-46,NOMA-Tree-47,NOMA-Tree-48,NOMA-Tree-49,NOMA-Tree-50,NOMA-Tree-51,NOMA-Tree-52,NOMA-Tree-53,NOMA-Tree-54,NOMA-Tree-55,NOMA-Tree-56,NOMA-Tree-57,NOMA-Tree-58,NOMA-Tree-59,NOMA-Tree-60,NOMA-Tree-61,NOMA-Tree-62,NOMA-Tree-63}

A taxonomy of NOMA schemes in RF communication systems is provided in Fig.~\ref{fig:NOMA_in_RF}. The figure illustrates the work available in the literature related to the three aforementioned versions of NOMA. In power-domain NOMA, the existing works are divided based on the system model setup that includes conventional NOMA, cooperative NOMA scheme, and NOMA scheme with cognitive radio (CR). The works on conventional power-domain NOMA are further classified based on network topology that may contain two users or multiple users, and SISO or MIMO transceivers in single-cell or multiple cells scenarios. Code-domain NOMA assigns different codes to different users to support multiple transmissions within the same time-frequency resource block~\cite{Modulation-and-Multiple-Access-for-5G-Networks-Survey}. Existing solutions for code-domain NOMA include LDS-CDMA, LDS-OFDM, SCMA, MUSA, and SAMA. NOMA multiplexing in multiple domains is proposed to support massive connectivity for 5G networks~\cite{Modulation-and-Multiple-Access-for-5G-Networks-Survey}. Existing solutions for NOMA multiplexing in multiple domains communication scheme include PDMA, BOMA, and LPMA.

\section{Single-user VLC Systems}
\label{section: Single-User Visible Light Communication Systems}

\begin{figure*}[!t]
\centering
\includegraphics[width=300pt]{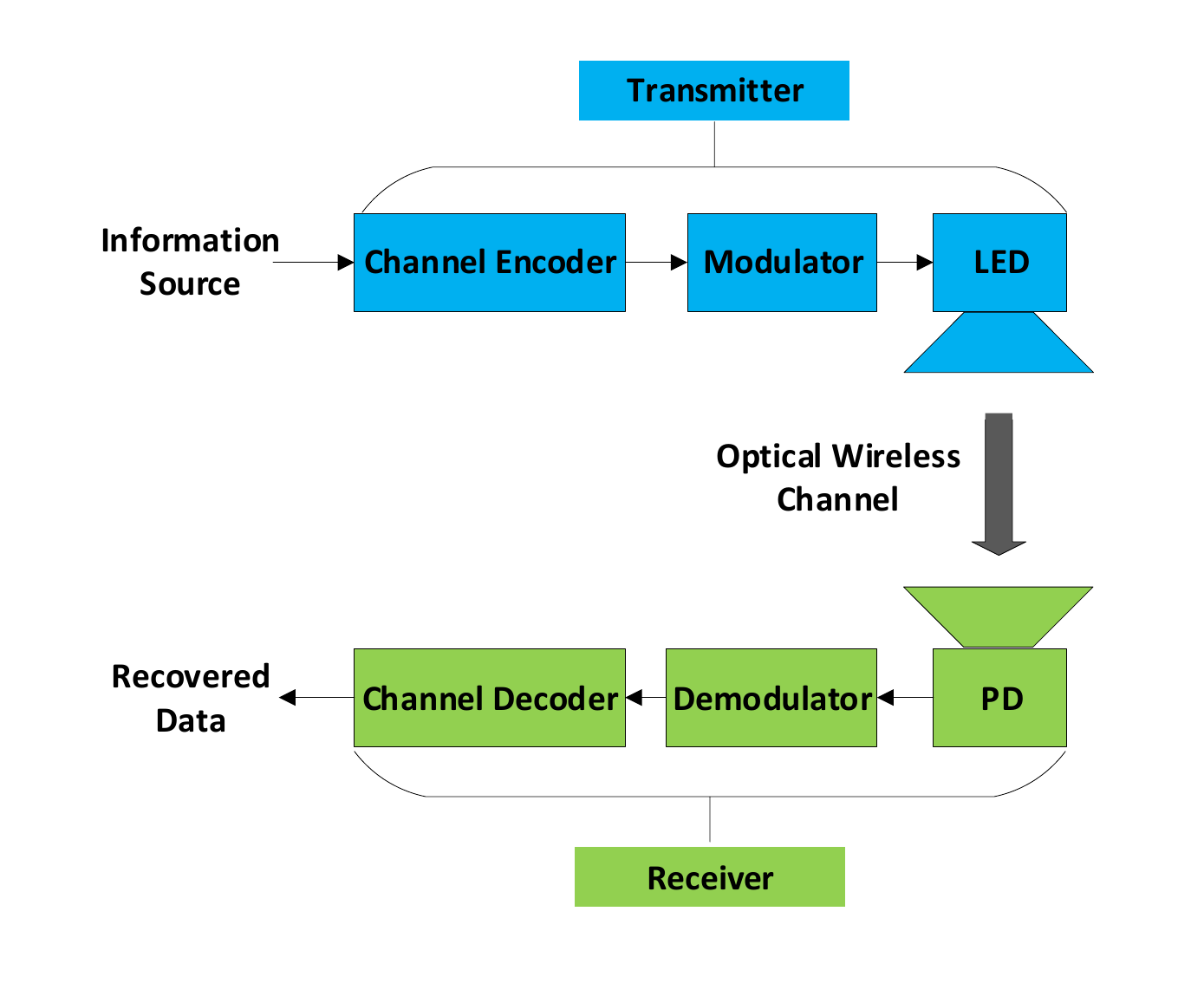}
\caption{A typical single-user VLC system block diagram~\cite{ow-surv-2}.}
\label{fig:VLC_P2P}
\end{figure*}

In this section, we present a brief overview of point-to-point VLC systems. In Fig.~\ref{fig:VLC_P2P}, a typical single-user VLC system is shown  where the transmitter typically consists of the channel encoder and the modulator followed by the optical front end. The electrical signal modulates the intensity of the optical carrier to send the information over the optical channel. At the receiver, a photodiode  receives the optical signal and converts into an electrical signal followed by the recovery of data.

\begin{figure*}[!t]
\centering
\includegraphics[width=500pt]{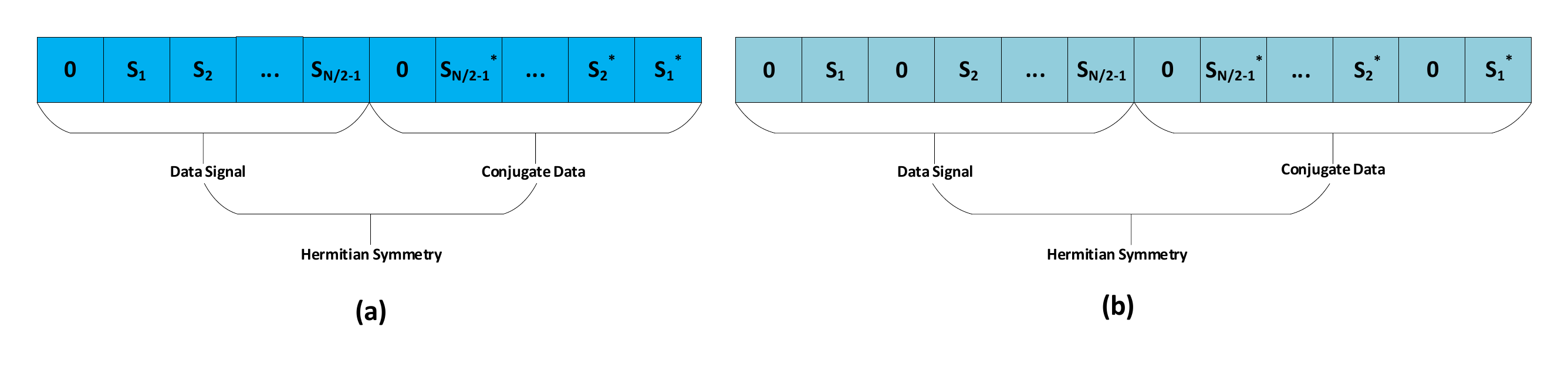}
\caption{Signalling structure for: (a) DCO-OFDM and (b) ACO-OFDM~\cite{vlc-noma-65,DCO-ACO-OFDM-Journal-2017}.}
\label{fig:Hermitian-symmetry-property}
\end{figure*}

In early VLC systems, single-carrier systems with intensity modulation techniques such as on-off keying (OOK), M-ary pulse-amplitude modulation (M-PAM), and M-ary pulse-position modulation (M-PPM) intensity modulation schemes were commonly used.
However, the demand for spectrally efficient high data-rate communication motivated the introduction of multiple-subcarrier modulation (MSM) schemes by Carruthers and Khan~\cite{optical-mc-1}. Later, optical OFDM~\cite{mu-mimo-book1} owing to its appealing performance over the dispersive communication channels and low implementation complexity was proposed. Since the modulating waveform has to be real-valued and non-negative in intensity modulation and direct detection (IM/DD) systems, several variants of optical OFDM were introduced in the literature~\cite{optical-ofdm-1,optical-ofdm-2,optical-ofdm-12,optical-ofdm-4, optical-ofdm-5}. Two most popular ones are DC-biased optical OFDM (DCO-OFDM), and asymmetrically clipped OFDM (ACO-OFDM). In these optical OFDM systems, Hermitian symmetry is imposed on data signal to obtain real valued signals as illustrated in Fig.~\ref{fig:Hermitian-symmetry-property}. To make the signals unipolar, a DC bias is added to shift the signal in DCO-OFDM. For the same purpose, only odd carriers are modulated in ACO-OFDM. This leads to a symmetry in time domain as illustrated in~[12, Fig.4.2]. Therefore, negative parts can be clipped without loss of information. Another issue in the implementation of OFDM in VLC systems is the non-linearity of the LEDs responses which further complicates the peak to average power ratio minimization.

The upcoming VLC standards IEEE 802.15.13 and IEEE 802.11.bb target peak data rates of 10 Gbit/sec and are expected to adopt DCO-OFDM. To reach such ambitious data rates, other advanced physical layer techniques such as adaptive transmission and MIMO techniques should be further adopted in conjunction with DCO-OFDM. Since multiple light sources typically exist in an indoor environment,  MIMO VLC systems provide a natural high-rate solution. In~\cite{MIMO-Techniques-in-OWC-trans}, Fath and Haas provided a comparative performance evaluation of three MIMO techniques, namely repetition code (RC), spatial multiplexing (SM) and spatial modulation (SMOD). In~\cite{MIMO-ACO-OFDM-system-Journal}, He et al. presented the performance analysis of a MIMO VLC system with SM using sub-optimal detection techniques such as ZF and MMSE. In~\cite{performance-evaluation-MIMO-VLC-Conf}, Damen et al. considered the combination of MIMO and OFDM and provided a performance comparison among RC, SM and SMOD techniques for multi-carrier VLC systems.

Adaptive transmission has been extensively studied in the context of RF communications and was recently applied to VLC systems. Adaptive transmission is based on the principle of selecting transmission parameters (e.g., modulation size, transmit power, code rate, etc.) according to the channel conditions. OFDM-based adaptive VLC systems were explored in~\cite{Adaptive-modulation-schemes-for-VLC-Journal,DMT-modulation-VLC-Journal,Rate-Adaptation-in-VLC-Conf} where bit and power loading are considered. In~\cite{Rate-Adaptation-in-VLC-Journal}, Wang et al. proposed a coded adaptive OFDM VLC system where code rate and modulation order are chosen as adaptive transmission parameters. In~\cite{Spatial-Multiplexing-in-Optical-MIMO-trans}, Park et al. considered an adaptive MIMO VLC system with SM technique and investigated power and bit loading. In~\cite{Adaptive-MIMO-OFDM-letter}, bit and power loading was studied for an OFDM MIMO VLC system. More recently, in~\cite{Link-Adaptation-MIMO-OFDM-Access}, Narmanlioglu et al. proposed an adaptive OFDM VLC system which allows MIMO mode switching based on channel conditions. Specifically, they devised a joint MIMO mode selection and bit loading scheme to maximize the spectral efficiency while satisfying a given bit error rate (BER) target.

VLC functionality should be provided as an add-on service of the luminary with primary illumination function. Therefore, flickering improvement and dimming techniques which are of critical importance in practical deployment attracted some attention in the literature~\cite{dimming-techniques-journal-1,dimming-techniques-journal-2,dimming-techniques-journal-3,dimming-techniques-letter-4,dimming-techniques-letter-5,dimming-techniques-letter-6}. Flicker refers to the fluctuation of the brightness of light. To quantify this, maximum flickering time period (MFTP) is defined. MFTP  is  the maximum time period over which the light intensity can change without the human eye able to perceive it. In  a VLC-enabled  luminary,  the  changes  in  brightness should  be  below the MFTP value. Dimming support is another practical  consideration for VLC systems. When the light source is dimmed,  communication  link  should  be maintained  albeit  at  lower  SNR  values.  A  comparison  of modulation techniques in terms of flickering mitigation and dimming support can be found in~\cite{Flickering-Dimming-comparison-Serv}.

\section{Multi-user VLC Systems}
\label{section: Multi-user Visible Light Communication Systems}

As  summarized  in  the  previous  section, the earlier literature on VLC has mainly focused on single-user VLC systems~\cite{vlc-book-2018, ow-surv-1, ow-surv-2,ow-surv-3,vlc-surv-1,vlc-surv-2,vlc-surv-3} with a  particular consideration of static and one-directional point-to-point links. While these  works demonstrate the feasibility of VLC concept, the widespread adoption can be only realized with a fully networked VLC system which supports multi-user access, mobility and bi-directional communication. To address such practical concerns, more recent research efforts have targeted the design, analysis and optimization of multi-user VLC systems. In the following, we first summarize the distinctive features of VLC that need to be taken into account in such designs. Then, we present the latest advances in multi-user VLC systems. In particular, we discuss the  existing  works on MU-MIMO, multiple-access schemes, resource allocation  and mobility management in the context of VLC highlighting major differences from RF systems described in the previous section.

\subsection{Key Design Considerations}

\begin{figure*}[!t]
\centering
\includegraphics[width=350pt]{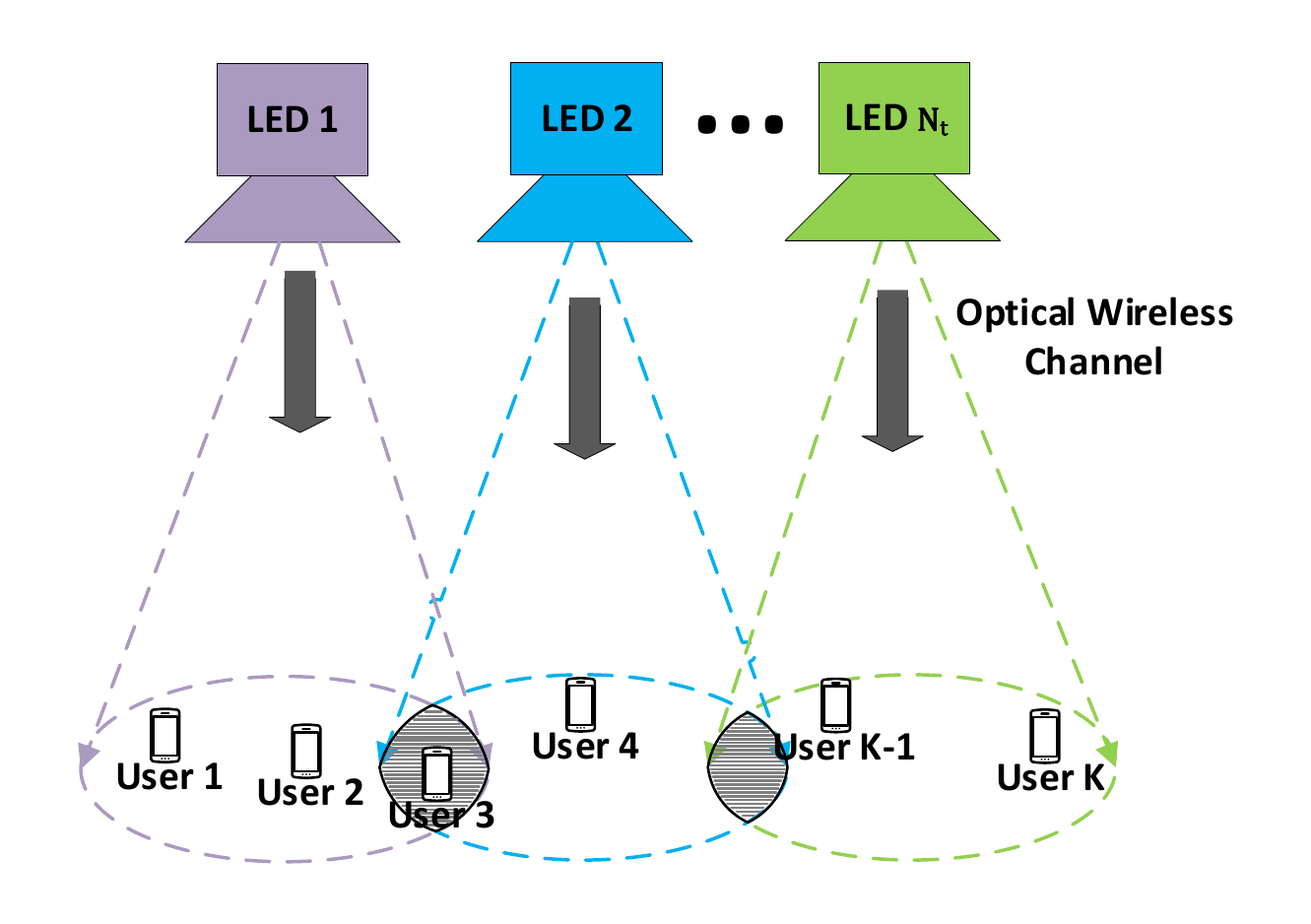}
\caption{A generic model for a multi-user VLC system~\cite{ow-surv-2}.}
\label{fig:VLC-multi-user}
\end{figure*}

A generic model for a multi-user VLC system is shown in Fig.~\ref{fig:VLC-multi-user} where a set of transmit $N_t$ LEDs communicate with a set of users. This is a typical indoor downlink scenario where the LEDs on the ceiling transmit the downlink data to the VLC receivers that are equipped with photodetectors~\cite{vlc-multiuser-1}. The field of view (FoV) of the LEDs and photodiodes are designed to satisfy both the illumination and communication requirements. Each shaded area  represents the ICI region imposed by the signal conveying different information and arriving from the neighbouring  cell. While there is no signal fading experienced in VLC channel, the signal-to-noise-interference ratio (SINR) significantly drops at the boundary of interference region and by obstruction of the line-of-sight (LOS) path due to user mobility. In a multi-user system, the transmission resources described in time, wavelength, and/or space are divided into resource units. A key challenge is the allocation of resource units in a way that key performance metrics such as user fairness, spectral efficiency, latency, energy efficiency, etc., are  fulfilled. The mobility of users further necessitates efficient handover techniques which eventually affect resource allocation and network stability. Such practical considerations have motivated research efforts on multi-user VLC systems. While most existing works on multi-user VLC systems are inspired from the RF literature, the distinctive features of VLC systems need to be considered. Some of these distinctive features are listed below with implications on system design:

\begin{itemize}
  \item The channel impulse response in indoor VLC systems consists typically of a LOS component and a non-LOS component (due to reflections from the room surfaces). The LOS component is commonly assumed to be dominant (even the strongest diffuse component is weak relative to the LOS components~\cite{vlc-models}) and the signal fading is negligible since almost no fading takes place in VLC channels due the large aperture size of the optical front-end as compared to the wavelength of the optical signal~\cite{vlc-models-1, vlc-models-2}. This is a main difference from RF channels where multipath fading is common. This quasi-static nature of VLC channels reduces the burden of the frequent estimation of CSI so that the implementation of CSI-dependent transmission scheme would be more feasible. Another implication of this is that opportunistic channel-aware scheduling is no more attractive in VLC systems.
  
  \item The coverage range is limited as the light waves can not penetrate through non-transparent objects such as walls and partitions. However, this feature offers security and efficient spatial re-use merits in contrast to RF systems. 
  
  \item  In addition to the requirement that the signals transmitted by LEDs must be non-negative and real-valued, the constraint of maximum permissible current of each transmitting LED (to avoid the non-linearity effects) has to be considered. This makes the design of the optimal and sub-optimal transmission schemes more difficult than the RF case especially for multi-user VLC communications as highlighted in the subsequent sections of this paper. In fact, the channel capacity even for single-user VLC channels is still an open problem and is known only for special cases~\cite{optical-capacity-1, optical-capacity-2, optical-capacity-3} . 
  
  \item The primary purpose of LEDs is illumination while VLC is provided as an add-on service. Therefore, both the illumination requirements and constraints such uniform illumination, fine-grained dimming control, and flickering avoidance, should be be considered in the design of VLC systems~\cite {joint-vlc-1,vlc-control-1}. For example, the design of a multi-user VLC would require highly directional LEDs to reduce the multi-user interference; however, such choice may not satisfy the illumination requirements and joint design might be needed.  
  
  \item Although the available bandwidth in the visible light band is abundant, the current off-the-shelf LEDs have a limited bandwidth, a few to tens of Megahertz, and the realization of high data-rate VLC has triggered both new optical enabling technologies~\cite{optical-design-1,optical-design-2,multi-user-vlc-wdma3, optical-design-3} and/or the adoption of "RF-inspired" advanced communication and signal processing schemes such as MIMO, MU-MIMO, spectrally efficient multiple access schemes, adaptive modulation and equalization, cooperative communication schemes such as relaying and CoMP, and others~\cite{optimal-vlc-1,joint-vlc-1, vlc-schemes}.

  \item The high correlations are common in VLC systems as the received signals in a LOS scenario are almost identical~\cite{MIMO-Techniques-in-OWC-trans}. Such correlations are expected to affect the performance of linear precoding schemes in VLC systems as highlighted later and some advanced receiver structures need to be used to reduce these high correlations~\cite{vlc-book-2018}.
  
  \item Uplink transmission using VLC leads to unpleasant irradiance from the user devices. So, the current technologies are expected to rely on RF or infrared links for uplink transmission.
  
\end{itemize}

\subsection{Precoding Schemes for Multi-user VLC Systems}

As discussed before, the capacity-achieving transmission schemes for VLC channels are generally unknown. Therefore, researchers have typically considered the schemes from RF literature and modified them for VLC systems. An early work on the use of multi-user MIMO schemes in VLC systems was conducted by Yu et al.~\cite{mu-mimo-vlc1} where the performance of linear ZF and ZF-DPC techniques, motivated by the high SNR in VLC systems, for a multi-user multi-input single-output (MU-MISO) downlink VLC system were compared through simulations. It was observed that ZF-DPC outperforms linear ZF in correlated scenarios as expected for ZF in ill-conditioned channels. In a later work on MISO downlink channels~\cite{mu-mimo-vlc2}, a set of $M$ LEDs transmit precoded data to $M$ users through the LOS channel. The precoding matrix and the receive coefficients at the users are optimized to minimize the minimum square error (MSE) between the transmitted and received signals of the users for the per-LED optical power constraints. Moreover, sub-optimal ZF-based transmit precoding strategy for per-LED optical power constraints was derived and it was shown that the two proposed schemes outperform the conventional pseudo-inverse ZF-based precoding in terms of the average MSE and the BER. A similar set-up was considered by Shen et al.~\cite{mu-mimo-vlc3} where the optimal ZF-based precoding matrix that maximizes the sum rate of the downlink users was obtained using the iterative concave-convex procedure. The optimization problem considered therein is more complex than its RF counterpart due to the constraint of maximal permissible current of each transmitting LED. Shen et al. also demonstrated that their proposed beamforming scheme outperforms the conventional pseudo-inverse  precoding in terms of the achievable users' sum rate. Similar results for an MU-MISO downlink VLC system were obtained in~\cite{mu-mimo-vlc31}. This is due to the fact that the pseudo-inverse option restricts the search for the optimal solution into a smaller feasible subset as compared to the optimal ZF-based precoding.   

The design of the linear precoding to maximize the sum rate without the ZF constraint was presented in~\cite{mu-mimo-vlc32}. The sequential parametric convex approximation (SPCA) method was utilized to obtain the optimal solution. The simulation results have shown that the proposed scheme outperforms the optimized ZF-based scheme in~\cite{mu-mimo-vlc3} for low LED optical power or high channel correlations. In~\cite{mu-mimo-vlc4}, the BD transmission scheme is implemented for a downlink VLC indoor system with two users. It was observed that the performance of the BD scheme, in terms of the BER, is limited by the high correlations at the receivers of each user. Consequently, a scheme based on the utilization of different FOV values on user equipment was proposed to improve the BER performance.    

In~\cite{mu-mimo-vlc7}, a multi-user OFDM MIMO indoor VLC system was considered where $N_t$ LEDs serve $N_r$ single-photodiode users. The multi-user interference is mitigated through transmit precoding per subcarriers using the ZF and MMSE schemes. The simulations have shown a significant increase in the achievable sum rate for the uncorrelated high SNR scenarios. Also, the bandwidth-power efficiency trade-off of both DCO-OFDM (with minimum DC bias and uniform DC bias) and ACO-OFDM was investigated. The use of linear precoding at both the transmitter and the receiver to reduce the inter-user interference was carried out in~\cite{mu-mimo-vlc-41}. The design of linear precoding to implement CoMP in VLC networks was considered in~\cite{mu-mimo-vlc5,mu-mimo-vlc52} using MSE and weighted  sum  mean  square  error (WSMSE)-based coordinated beamforing as further discussed in Section IV. C. The aforementioned works on the precoding schemes for VLC systems and networks are summarized in Table 1.        
 
\begin{table*}[t]
\caption{ \textbf{A summary of precoding schemes for VLC systems.}}
\centering 
\begin{tabular}{ |p{0.15\linewidth}|p{0.3\linewidth}|p{0.45\linewidth}| } \hline

\ \ \ \ \ \ \  System & 
\ \ \ \ \ \ \ \ \ \ Precoding scheme &
\ \ \ \ \ \ \ \ \ \ \ \ \ \ \ \ \ \ \ \ \ \ \ \ Comments \\\hline

Multi-user MISO & ZF-based linear precoding to minimize MSE~\cite{mu-mimo-vlc2} &  The proposed scheme outperforms conventional pseudo-inverse ZF-based precoding; however, performance is limited by receivers correlations. \\\hline

Multi-user MISO &  ZF-based linear precoding~\cite{mu-mimo-vlc3, mu-mimo-vlc31}   and linear preoding without the ZF constraint~\cite{mu-mimo-vlc32} to maximize the sum rate   &  The proposed schemes outperform conventional pseudo-inverse ZF-based precoding; however, the scheme in~\cite{mu-mimo-vlc32} outperforms the one~\cite{mu-mimo-vlc3} for high channel correlations.  \\\hline

Multi-user OFDM-MISO \cite{mu-mimo-vlc7} & Conventional pseudo-inverse ZF-based and MMSE & The performance is very limited by the users' correlations. \\\hline

multi-user CoMP-MISO &  Linear precoding to minimize the sum-MSE of the users~\cite{mu-mimo-vlc5} or the WSMSE~\cite{mu-mimo-vlc52} & Significant SINR gain due to coordination; however, the SINR may saturate with partial cooperation.  \\\hline

Multi-user MIMO ~\cite{mu-mimo-vlc-41} & Linear precoding at both the transmitter and receiver for the max-min SINR optimization & The proposed MU-MIMO system has a significant gain compared to the MU-MISO case. \\\hline

Multi-user MIMO \cite{mu-mimo-vlc4} &  Block diagonalization (BD) precoding  &  The performance is limited by receive correlations per user.\\\hline

\end{tabular}
\label{table1}
\end{table*}

\subsection{Multiple Access Schemes for VLC Systems}
\label{subsection:NOMA for VLC Systems}

\begin{figure*}[!t]
\centering
\includegraphics[width=350pt]{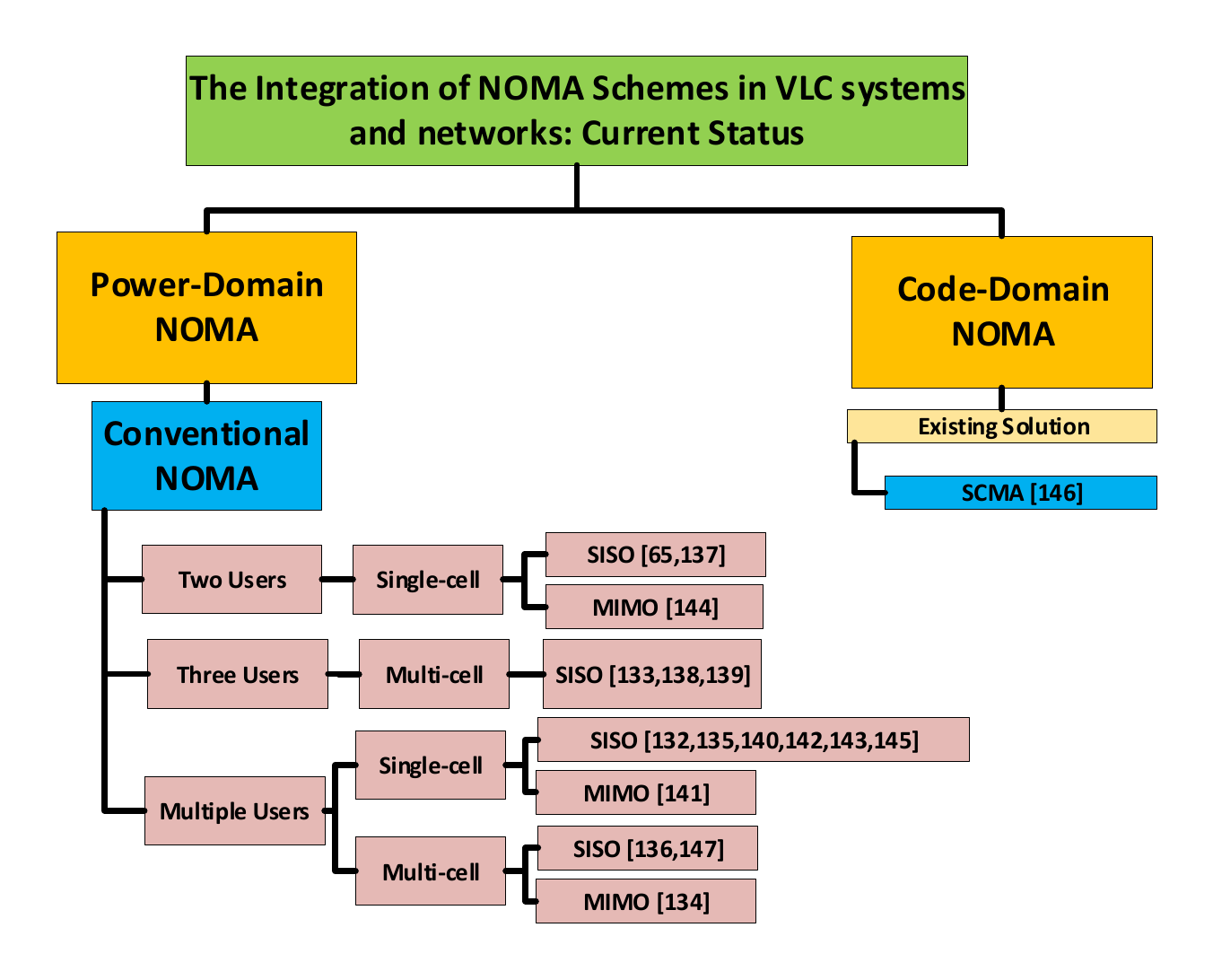}
\caption{A summary of the integration of NOMA schemes in VLC systems and networks.}
\label{fig:NOMA-VLC-tree}
\end{figure*}

The conventional orthogonal multiple access schemes used in VLC systems are TDMA (time division multiple access), OCDMA (optical code division multiple access), WDMA (wavelength division multiple access), and SDMA (space-division multiple access). In TDMA, the different users are assigned different time slots in an orthogonal manner that removes interference due to the overlap of the coverage area of the transmitting  laser diodes (LDs) or the LEDs~\cite{multi-user-vlc-tdma1}. However, TDMA is not an attractive option for multi-user VLC systems due to synchronization and transmit power limitations between the transmitter and the receiver~\cite{multi-user-vlc-tdma2, vlc-models-2,multi-user-vlc-tdma4}. OCDMA is based on the direct-sequence spread spectrum technique in a coherent or non-coherent manner~\cite{optical-MA-book1}. The commonly used spreading codes are the one- dimensional codes such as Gold sequences, optical orthogonal codes and prime codes and the two-dimensional orthogonal optical codes. The two-dimensional codes spread in both time and wavelength domain. OCDMA-based VLC systems were investigated in~\cite{multi-user-vlc-ocdma1,multi-user-vlc-ocdma2, multi-user-vlc-ocdma3,multi-user-vlc-ocdma4}. 

In WDMA, the users transmit simultaneously using different non-interfering wavelengths, using multicolor LEDs  \cite{multi-user-vlc-wdma1, multi-user-vlc-wdma2, multi-user-vlc-wdma3}, while in SDMA, multiple directional beams at the transmitter are utilized to serve a set of narrow field-of-view receivers to reduce path loss, delay spead, and inter-user interference  \cite{multi-user-vlc-sdma1, multi-user-vlc-sdma2, multi-user-vlc-sdma3}. However both schemes are relatively complex to implement plus the difficulty in realizing dense WDMA using the current off-the-shelf LEDs \cite{multi-user-vlc-sdma2, multi-user-vlc-sdma4}.  

The other common multiple access scheme that was adopted in VLC systems is optical OFDMA \cite{vlc-ofdma-1}. The early work in \cite{vlc-ofdma-2} has considered the discrete multi-tone (DMT) based VLC system where a heuristic algorithm for the subcarriers and allocation was proposed and the performance gain as compared with a conventional DMT scheme was demonstrated. Later, the joint design of the DC-bias level, power and sub-carrier allocation in a DCO-OFDMA VLC system was considered in \cite{vlc-ofdma-4} where algorithms for the bias optimization (for a given power and subcarrier allocation), power and subcarrier allocation, and the joint optimization were proposed and simulations to compare the achievable sum rates of the different algorithms were carried out.      

NOMA is a recent promising multiple access scheme which was introduced in RF literature as discussed in Subsection~\ref{subsection: NOMA}. The adoption of NOMA in VLC systems was motivated by the following: (i) the VLC systems are used for off-loading in indoor environments, (ii) NOMA is favorable for high data-rate VLC systems as the current off-the-shelf LEDs have limited bandwidth, (iii) the quasi-static nature of the propagation channel lets reliable estimation of the channel gains for subsequent power allocation, and, (iv) under typical illumination constraints, VLC experiences high SNR conditions where it well known that NOMA schemes outperform the orthogonal counterparts in that particular region. This has motivated the study of NOMA in the context of VLC systems~\cite{vlc-noma-65,vlc-noma-134}. For example, performance optimization of NOMA in multi-user VLC systems has been carried out by Yin et al.~\cite{vlc-noma-134}. The proposed downlink VLC system maximizes the system coverage probability by finding the optimum power allocation coefficients set. Analytical results show that the capacity of NOMA VLC system outperforms the capacity of VLC-TDMA system by choosing LEDs with proper semi-angle. Thereafter, in~\cite{vlc-noma-65}, Kizilirmak et al. have proposed a novel DCO-OFDM based power-domain NOMA scheme. The numerical results in their work have demonstrated the capacity of the NOMA scheme and quantified the effect of the cancellation error in the SIC receiver compered with OFDMA scheme.  

Most of the existing works on NOMA schemes in multi-user VLC networks have addressed the development of power allocation (PA) and ICI mitigation techniques. For the PA techniques, researchers have investigated the fixed PA~\cite{vlc-noma-65}, the exhaustive search PA~\cite{vlc-noma-134}, the gain ratio PA~\cite{vlc-noma-135}, the normalized gain difference PA~\cite{vlc-noma-136}, the sum rate maximization PA~\cite{vlc-noma-137}, the sum rate maximization under quality of service (QoS) constraints PA~\cite{vlc-noma-138}, and the max-min fairness with sum rate maximization PA~\cite{vlc-noma-139}. While for the ICI mitigation techniques, researchers have proposed cell zooming technique~\cite{vlc-noma-135}, location-based user grouping technique~\cite{vlc-noma-138}, and a technique for handling the users in the overlapping VLC cells~\cite{vlc-noma-140}. In addition, research has been done on the BER and symbol error rate (SER) performance analysis by Huang et al., Marshoud et al., Mitra and Bhatia~\cite{vlc-noma-141,vlc-noma-142,vlc-noma-143}.

\nocite{vlc-noma-144,vlc-noma-145,vlc-noma-146,vlc-noma-147,vlc-noma-148,vlc-noma-149}

A summary of NOMA schemes in VLC literature is provided in Fig.~\ref{fig:NOMA-VLC-tree}. It is obvious that, similar to RF systems, most of the research has focused on power-domain NOMA and only one reference has investigated the SCMA scheme in code-domain NOMA in VLC systems. In addition, the SISO case is assumed for all two-users/multiple-users single-cell/multi-cell scenarios due to its simplicity in the analysis while there is still limited amount of research on the more involved MIMO case. Moreover, in multi-cell NOMA scenarios, the additional ICI mitigation and mobility management requirements were addressed since these requirements become a major obstacle in achieving the benefits of NOMA~\cite{NOMA-Tree-40}.

\subsection{Resource Allocation and Mobility Management in VLC Networks}

The research work on VLC, beyond the single-cell downlink scenarios, has led to two proposals for the integration of VLC in wireless networks, which are:  (i) hybrid or heterogeneous RF/VLC~\cite{vlc-rf-networks-1,vlc-rf-networks-2,vlc-rf-networks-3}, and, (ii) VLC-based visible light networks (VLNs)~\cite{vlc-networks-1,vlc-networks-2}. This motivated the research on interference management, resource allocation, mobility management, and multi-user scheduling techniques for these networks. This section is intended to overview these techniques from the physical layer perspective.   

Similar to the concept of CoMP in RF systems~\cite{comp-1, comp-2}, where a set of cells coordinate their transmissions to reduce the network interference, the interference in VLC attocells (where each LED acts as an optical AP to serve multiple users within its coverage)~\cite{vlc-surv-4} which occurs due to uniform illumination of the LEDs is accommodated through the introduction of the cooperation among multiple VLC attocells using power line communications~\cite{mu-mimo-vlc5}. In this scheme, $N_t$ LEDs cooperate to broadcast information to $N_r$ single-photodiode users using linear precoding that is optimized to minimize the sum-MSE under illumination constraints or to minimize the illumination level for a given set of MSE thresholds for all users. The first optimization problem, to minimize the sum-MSE, is solved using iterative optimization as proposed by Li et al.~\cite{mu-mimo-vlc2}. While the second was solved using semi-definite relaxation method. They demonstrated via numerical results that there is a significant SINR increase when full coordination is implemented. Further work~\cite{mu-mimo-vlc52}, by the same authors, has considered WSMSE-based coordinated beamforming to minimize the inter-attocell interference where only the downlink CSI is exchanged among the cooperating attocells as compared to the joint transmission scheme proposed by Ma et al.~\cite{mu-mimo-vlc5}. It was shown that coordinated beamforming has a comparable performance to joint transmission for some user distributions. There are three additional approaches to reduce the network interference in VLC networks. The first approach depends on fractional frequency reuse concept~\cite{vlc-rf-networks-3,interference-mitigation2-Journal}, the second approach provides a dynamic scheduling for VLC network resources based on users location~\cite{interference-mitigation3-letter,interference-mitigation4-Journal,vlc-networks-2,interference-mitigation6-Journal}, and the third approach achieves a dynamic pairing between the VLC users and the VLC APs~\cite{interference-mitigation7-Journal,interference-mitigation8-Journal,interference-mitigation1-Journal}.  

Early work on resource allocation in VLC-RF networks by Rahaim et al.~\cite{vlc-rf-networks-1} considered an indoor combined VLC attocell and RF femtocell system. They proposed decentralized algorithms for optimal resource allocation for different types of mobile terminals. Later, Kashef et al.~\cite{vlc-rf-networks-energy-1} investigated the optimization of bandwidth and power allocation to maximize the energy efficiency of heterogeneous VLC-RF networks and has shown that hybrid VLC-RF networks outperform RF-only networks for practical ranges of transmit power. The spectrum and power allocation for an RF system that complements a VLC system to improve coverage for randomly distributed users was considered by Basnayaka and Hass~\cite{vlc-rf-networks-rates-1}. They demonstrated that the per user rate outage probability improves as dynamic resource allocation is adopted.

Mobility management is another important issue in VLC networks where the handover process needs to be optimally carried out. There are two types of handover schemes in VLC networks, namely: vertical  (inter-system) handover and horizontal (intra-system) handover~\cite{vertical-Handover-serv}. The former takes place between different access technologies such as the vertical handover in the hybrid RF/VLC networks, while the latter allows the mobile user to transfer from one LED to another seamlessly while moving. Vertical handover (VHO) is usually invoked in order to maintain a smooth QoS for users in hybrid RF/VLC systems. VHO procedures occur either when an RF user receives a stronger VLC signal than its own RF signal, or when a VLC user loses the LOS transmission with the VLC AP or it moves to the cell edge~\cite{vlc-surv-2}. Rahaim et al.~\cite{vlc-rf-networks-1} proposed a VHO criteria for a hybrid RF/VLC systems that integrates an omnidirectional RF channel with directional broadcast VLC channels. The proposed criteria improves both the total throughput and the service quality of the system. Later, Bao et al.~\cite{VHO-Journal-2} proposed a hybrid OFDMA/VLC system which includes VLC APs for downlink transmission only and one OFDMA AP for both uplink and downlink transmission. The proposed system supports the horizontal and the vertical handover mechanisms for mobile users and achieves large improvements in the capacity performance as compared to an equivalent OFDMA system.

In a typical VLC network, the "cell" covers only a few square meters~\cite{study-2017}, therefore user movement may prompt very  frequent handovers that are not typical in RF-based networks. To handle such frequent handovers, both soft and hard handover strategies were studied in the context of VLC networks~\cite{soft-hard-handover-book}. In the soft handover, the connection to the neighbour LED is made before breaking the connection with the current LED which is known as a "make-before-break" method. While in the hard handover, the connection to the neighbour LED is made after breaking the connection with the current LED which is known as "break-before-make" method. The work on horizontal handover in the context of VLC networks are reported in~\cite{Horizontal-handover-vlc-networks1-Conf,Horizontal-handover-vlc-networks3-Conf, Horizontal-handover-vlc-networks5-letter, Horizontal-handover-vlc-networks7-Conf} where there are four main proposed approaches: (i) the first approach exploits the analogy between the received signal strength (RSS) parameter in the RF domain and the received signal intensity (RSI) parameter in the optical domain in order to develop an RSI-based handover mechanism for mobile VLC users~\cite{Horizontal-handover-vlc-networks1-Conf},  (ii) the second approach depends on changing the coverage region of the VLC cells dynamically while maintaining the desired illuminance in the environment~\cite{Horizontal-handover-vlc-networks3-Conf},  (iii) the third approach relies on finding an optimal LED footprint mapping while considering various practical parameters such as the number of users in the network, the number of handovers, and user mobility~\cite{Horizontal-handover-vlc-networks5-letter}, and finally, (iv) the fourth approach is based on the CoMP transmission scheme~\cite{Horizontal-handover-vlc-networks7-Conf}.

While the horizontal handover is mainly deployed in VLC networks to reduce the effect of ICI, multi-user scheduling is of comparable importance to reduce the effect of the inter-user interference. Tao et al.~\cite{interference-mitigation4-Journal} proposed a low complexity multi-user scheduling scheme to coordinate the inter-user interference for indoor multi-user VLC systems. The proposed scheme aims to maximize the sum capacity of the system while taking into account user fairness. A novel anticipatory design based on predicted users' future locations and their predicted traffic dynamics has been proposed by Zhang et al.~\cite{user-association-1-trans2018-hass-hanzo} as a beneficial design for user-to-AP associations. The proposed design provides an insight on the performance trade-off between the average system delay and the average per-user throughput in dynamic indoor VLC networks.

\section{Conclusion and Directions for Future Research}
\label{section: Conclusion and Directions for Future Research}

The emergence of VLC systems as high data-rate, secure, and energy efficient downlink transmission technology has fueled research on the enabling communication schemes for these systems. In this paper, after a review of some background on the multi-user communications in RF systems, we highlighted the main distinctive features of VLC systems and discussed the state-of-the-art integration of the emerging MU-MIMO, non-orthogonal multiple access, and interference mitigation multi-user schemes in VLC systems. In the following, we highlight some of the possible directions of future research on the physical-layer based multi-user schemes for VLC systems. 

\subsection{MU-MIMO Precoding Schemes}

The current work on precoding schemes for VLC systems have mainly built on the known linear precoding schemes in the RF literature. However, the optimal linear precoding schemes for a general MU-MIMO VLC system for the different performance metrics such as the max-min rate and the maximum sum rate are still open problems for research. On the other hand, the design and performance of non-linear precoding schemes such as vector perturbation~\cite{mu-mimo-nonlinear-1, mu-mimo-nonlinear-2} and Tomlinson-Harashima precoding~\cite{mu-mimo-nonlinear-3, mu-mimo-nonlinear-4} might be of interest to enhance the spectral efficiency and to reduce the effect of the high channel correlations, being more common in VLC systems as compared to RF systems, which tend to degrade the performance of the linear precoding schemes. A recent work by Tagliaferri et al.~\cite{mu-mimo-nonlinear-vlc-1} has already considered the use of Tomlinson-Harashima precoding to mitigate the multi-user interference in a multi-user MISO VLC system. These open research problems can be also extended to massive MIMO VLC systems (i.e., large arrays of transmitting LEDs and/or receiving photodiodes)~\cite{large-scale-mimo-1,large-scale-mimo-2}. A recent work by Jain et al.~\cite{vlc-large-mimo-1} has proposed a singular value decomposition-based precoding scheme to improve the BER that is degraded by the high condition number of the channel matrix.         

Another interesting line of research is the utilization of the interference alignment (IA) technique. IA was originally proposed as a high-SNR optimal transmission scheme for interference channels. In IA, each sender encodes its signals over multiple dimensions of the signal space to align interference from other transmitters in the least possible dimension subspace at each receiver and provide the remaining dimensions for the desired signal~\cite{interference-1}. Later, this technique was adopted for other wireless systems and networks~\cite{interference-2}. In~\cite{vlc-interference-1}, the blind interference alignment (BIA) scheme, which relaxes the full CSI requirement at the transmitter~\cite{interference-3}, was considered for an MU-MISO indoor DCO-OFDM based VLC system where it was shown that it outperforms orthogonal multiple access schemes in terms of spectral efficiency and achievable BER. The implementation of BIA using a re-configurable photodetector at each user receiver was investigated in~\cite{vlc-interference-2}.  This choice is motivated by the less stringent CSI availability, transmit cooperation, and channel correlation constraints as compared to conventional linear precoding schemes.

\subsection{Further Issues on NOMA VLC}

As discussed in Subsection~\ref{subsection:NOMA for VLC Systems}, the adoption of NOMA schemes in VLC systems and networks has attracted a lot of research. Nevertheless, this area of research is still in its infancy as different aspects of the NOMA schemes are not yet well understood in VLC for both the single-cell and multi-cell scenarios. In the following, we highlight some challenges and open research problems that need to be addressed in the future as: i) VLC networks suffer from several challenges, such as the broadcast nature of the illumination sources, the limited VLC coverage within an opaque space, and the fact that VLC receiver should be facing the VLC AP. In order to overcome these limitations, combined use of VLC and RF networks is typically preferred where VLC is used for data off-loading purposes as a complementary technology. While there is a growing literature on hybrid RF/VLC networks, hybrid RF/VLC NOMA schemes are not yet studied. ii) While downlink has been intensively studied, uplink NOMA VLC needs further investigation. As of now, only one study~\cite{vlc-noma-145} has addressed the BER performance in the uplink VLC based NOMA-OFDMA system. iii) Relay-assisted VLC systems were introduced in the literature~\cite{Relay-Assisted-VLC-trans} where secondary lights (such as task lights, desk lamps) assist the ceiling lights acting as major data sources. Relay-assisted NOMA VLC is worthy of study particularly for improved coverage of cell-edge users. iv) As mentioned in Subsection~\ref{subsection: NOMA}, there are three major versions of NOMA, namely, power-domain NOMA, code-domain NOMA, and NOMA multiplexing in multiple domains. As seen in Fig.~\ref{fig:NOMA-VLC-tree}, most of the research has focused on power-domain NOMA in VLC systems and only one reference~\cite{vlc-noma-148} has investigated SCMA scheme in code-domain NOMA. Hence, further detailed investigations on adopting other code-domain NOMA solutions and the solutions of NOMA multiplexing in multiple domains (see Fig.~\ref{fig:NOMA_in_RF}) might be of interest to increase the capacity of NOMA VLC systems. (v) The adoption of NOMA will introduce some intra-cell interference that may increase with SIC decoding errors and affect both the downlink and uplink data transmissions. So, more sophisticated power and resource allocation schemes are expected in multi-cell scenarios with NOMA and some form of hybrid OMA-NOMA is expected to optimize the network performance. (vi) VLC channels tend to be correlated, as highlighted before in Section IV, and the role and effect of such correlations on the beamforming in NOMA VLC systems need to be investigated.      

\subsection{Optimized Resource Allocation for Multi-User VLC Networks}

In a conventional VLC network, each LED acts as an access point and the LEDs are connected to each other through electrical grid and data backbone. These VLC-enabled LEDs consist of baseband unit (BBU) followed by the optical front-end. In a recent work~\cite{C-LiAN-IEEE-access}, 
so-called “Centralized Light Access Network (C-LiAN)” was proposed that aggregates all BBU computational resources into a central pool that is managed by a centralized controller. Unlike the distributed architecture where each LED is supposed to perform both baseband processing and optical transmission/reception, "dummy" LEDs (with only optical front-ends) can be employed in the centralized architecture. Such an architecture enables joint processing of signals from different access points. In such centralized networks where the BBUs are physically/virtually are placed close to each other and can easily share the CSI and other signalling information, investigation of enhanced coordinated transmission techniques and handover management techniques remain as open research problems.

\subsection{Cognitive Radio}

The cognitive radio technology where the unlicensed (secondary) users (SUs) are allowed to share the spectrum with the licensed (primary) users (PUs) according to a certain protocol has attracted research in wireless systems for better spectrum utilization. In a recent work~\cite{vlc-cognitive-1}, an OFDMA-based VLC network is considered where the coverage area of each access point is divided into two zones serving the primary and secondary users, respectively. The maximization of the area spectral efficiency while satisfying the illumination, mobility, and handover requirements was carried out through the proper zone radius and the sub-carrier allocation. Another CR-motivated multi-cell VLC system was considered in~\cite{vlc-cognitive-2} where the users with stringent delay constraints and service requirements are set to be the PUs while the users with less stringent requirements are set to be the SUs. The SUs share the subchannels of PUs through the overlay and underlay spectrum access protocols using dynamic subchannel and power allocation. The sum rate improvement as compared to non-cognitive multi-cell VLC systems was demonstrated. Future work may consider the design and performance of CR schemes in VLNs (VLC-based networks) and RF/VLC networks as well as relay-enabled CR schemes for these networks for the MISO and MIMO scenarios.

\subsection{Physical Layer Security}

Physical layer security has recently emerged as an additional less computationally-intensive and flexible level of security compared to the existing security measures at the other layers in RF networks~\cite{security-1}. This has motivated the research in the relatively more secure but yet still vulnerable to security threats VLC channels. However, VLC channels are mainly amplitude-constrained channels whose channel capacity and secrecy capacity are still open problems~\cite{optical-capacity-1, security-2, security-3}, and only lower and upper bounds on the capacity of SISO Gaussian wiretap channels were derived. A recent work on the multi-user scenario appeared in~\cite{mu-vlc-security-1} where lower bounds on the average secrecy sum rate of the users for both cases of known and unknown eavesdroppers' CSI at the transmitter for ZF precoding were derived. Clearly, there are open research problems on the physical layer security for both VLC-based and hybrid RF/VLC multi-user systems (with different security requirements) for the different precoding schemes and under different CSI availability constraints. In addition to MIMO precoding, cooperative anti-jamming and relay-assisted security techniques are also of interest in VLN and hybrid RF/VLC networks. Another important topic is the security of NOMA VLC systems in the presence of both single and multiple eavesdroppers \cite{vlc-noma-198} and for the various network architectures.

\section*{Acknowledgments}
Authors acknowledge King Fahd University of Petroleum \& Minerals, Dhahran, Saudi Arabia, for all support. The work of M. Uysal was also supported by the Turkish Scientific and Research Council (TUBITAK) under Grant 215E311.

\bibliographystyle{IEEEtran}
\bibliography{IEEEabrv,main}

\vskip -2\baselineskip plus -1fil

\begin{IEEEbiography}[{\includegraphics[width=1in,height=1.25in,clip,keepaspectratio]{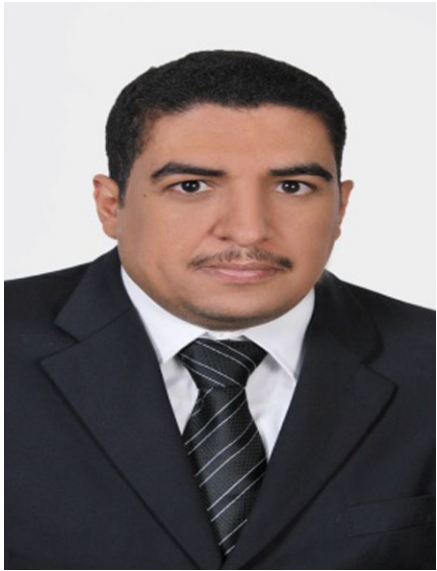}}]{\textbf{Saad Al-Ahmadi}}
has received his M.Sc. in Electrical Engineering from King Fahd University of Petroleum \& Minerals (KFUPM), Dhahran, Saudi Arabia, in 2002 and his Ph.D. in Electrical and Computer Engineering from Ottawa-Carleton Institute for ECE (OCIECE), Ottawa, Canada, in 2010. He is currently with the Department of Electrical Engineering at KFUPM as an Associate Professor. His current research interests include channel characterization, design, and performance analysis of wireless communication systems.
\end{IEEEbiography}

\begin{IEEEbiography}[{\includegraphics[width=1in,height=1.25in,clip,keepaspectratio]{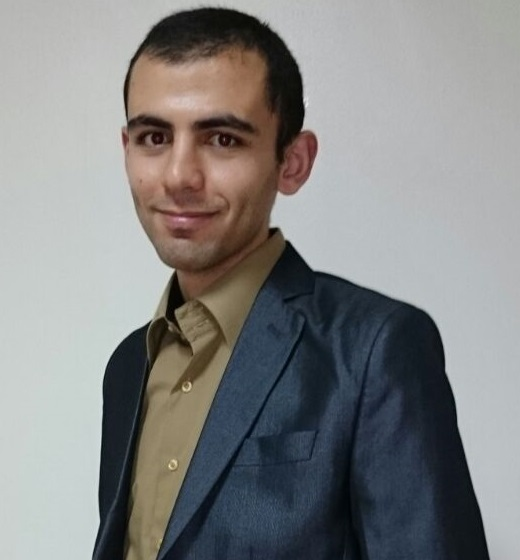}}]{\textbf{Omar Maraqa}}
has received his B.S. degree in Electrical Engineering from Palestine Polytechnic University, Palestine, in 2011, and his M.S. degree in Computer Engineering from King Fahd University of Petroleum \& Minerals (KFUPM), Dhahran, Saudi Arabia, in 2016. He is currently pursuing a Ph.D. degree in Electrical Engineering at KFUPM, Dhahran, Saudi Arabia. His research interests include non-orthogonal multiple access (NOMA), visible light communications, resource allocation, and optimization.
\end{IEEEbiography}

\begin{IEEEbiography}[{\includegraphics[width=1in,height=1.25in,clip,keepaspectratio]{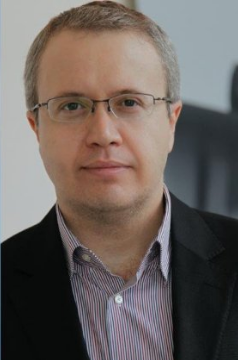}}]{\textbf{Murat Uysal}}
(S’98–M’02–SM’07) received the B.Sc. and M.Sc. degree in electronics and communication engineering from Istanbul Technical University, Istanbul, Turkey, in 1995 and 1998, respectively, and the Ph.D. degree in electrical engineering from Texas A\&M University, College Station, Texas, in 2001. He is currently a Full Professor and a Chair of the Department of Electrical and Electronics Engineering, Özyeğin University, Istanbul, Turkey. He also serves as the Founding Director of the Center of Excellence in Optical Wireless Communicaion Technologies. Prior to joining Özyeğin University, he was a tenured Associate Professor with the University of Waterloo, Canada, where he still holds an Adjunct Faculty position. His research interests are in the broad areas of communication theory and signal processing with a particular emphasis on the physical layer aspects of wireless communication systems in radio and optical frequency bands. Prof. Uysal was a recipient of the Marsland Faculty Fellowship in 2004, NSERC Discovery Accelerator Supplement Award in 2008, the University of Waterloo Engineering Research Excellence Award in 2010, Turkish Academy of Sciences Distinguished Young Scientist Award in 2011, and Özyeğin University Best Researcher Award in 2014 among others. He served as the Chair of the Communication Theory Symposium of the IEEE ICC 2007, a Chair of the Communications and Networking Symposium of the IEEE CCECE 2008, a Chair of the Communication and Information Theory Symposium of IWCMC 2011, a TPC Co-Chair of the IEEE WCNC 2014, and a General Chair of the IEEE IWOW 2015. He currently serves on the editorial board of the IEEE TRANSACTIONS ON WIRELESS COMMUNICATIONS. In the past, he was an Editor for the IEEE TRANSACTIONS ON COMMUNICATIONS, the IEEE TRANSACTIONS ON VEHICULAR TECHNOLOGY, the IEEE COMMUNICATIONS LETTERS, the Wiley Wireless Communications and Mobile Computing Journal, Wiley Transactions on Emerging Telecommunications Technologies, and a Guest Editor of the IEEE JSAC Special Issues on Optical Wireless Communication in 2009 and 2015. He was involved in the organization of several IEEE conferences at various levels. Over the years, he has served on the technical program committee of over 100 international conferences and workshops in the communications area.
\end{IEEEbiography}

\begin{IEEEbiography}[{\includegraphics[width=1in,height=1.25in,clip,keepaspectratio]{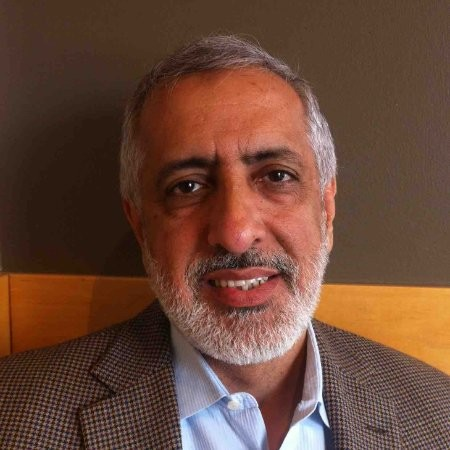}}]{\textbf{Sadiq M. Sait}}
(SM’02) was born in Bengaluru. He received the bachelor’s degree in electronics engineering from Bangalore University in 1981, and the master’s and Ph.D. degrees in electrical engineering from the King Fahd University of Petroleum \& Minerals (KFUPM) in 1983 and 1987, respectively. He is currently a Professor of Computer Engineering and the Director of the Center for Communications and IT Research, Research Institute, KFUPM. He has authored over 200 research papers, contributed chapters to technical books, and lectured in over 25 countries. He is also the Principle Author of two books. He received the Best Electronic Engineer Award from the Indian Institute of Electrical Engineers, Bengaluru, in 1981.
\end{IEEEbiography}

\end{document}